\documentclass{article}

\usepackage{arxiv}

\usepackage[utf8]{inputenc} 
\usepackage[T1]{fontenc}    
\usepackage{hyperref}       
\usepackage{url}            
\usepackage{booktabs}       
\usepackage{amsfonts}       
\usepackage{nicefrac}       
\usepackage{microtype}      
\usepackage{lipsum}
\usepackage{graphicx}
\usepackage{amsmath}
\usepackage{algorithm}
\usepackage{algpseudocode}
\usepackage[absolute,overlay]{textpos}
\graphicspath{ {./images/} }

\title{U-PAST: A Phase-Aware Audio Spectrogram Transformer-U-Net for Single-Channel Speech Enhancement}

\author{
 Cao Duong Ly \\
  Department of Medical Physics and Acoustics\\
  Carl von Ossietzky University Oldenburg\\
  Oldenburg, Germany 26129 \\
  \texttt{cao.duong.ly@uni-oldenburg.de} \\
   \And
 Jörn Anemüller \\
  Department of Medical Physics and Acoustics\\
  Carl von Ossietzky University Oldenburg\\
  Oldenburg, Germany 26129 \\
  \texttt{joern.anemueller@uni-oldenburg.de} \\
}

\begin{document}

\begin{textblock*}{12cm}(7cm,26cm)
\raggedleft
{\footnotesize
Model and live demonstration are available on Hugging Face:
\url{https://huggingface.co/spaces/lycaoduong/U-Past}
}
\end{textblock*}

\maketitle
\begin{abstract}
Convolutional neural networks (CNNs), used widely and successfully in audio enhancement, capture long-range time-frequency dependencies only indirectly, through successive convolution and pooling. 
Here, we present U-PAST, a hybrid transformer-U-Net architecture that addresses this limitation through self-attention dependency-modeling in the complex spectrogram domain.
U-PAST tokenizes a complex STFT representation, similarly to the magnitude spectrogram tokenization of the Audio Spectrogram Transformer (AST), applies a multi-layer transformer encoder, and reconstructs the enhanced complex spectrogram with a U-Net-style decoder.

We evaluate four architectural variants with between 1.17M and 2.40M parameters on the DNS Challenge, VoiceBank-DEMAND, and LibriMix corpora under matched, acoustic mismatch, and two-dataset mismatch conditions.
U-PAST attains the best SI-SDR of any evaluated model under acoustic mismatch and closely trails substantially larger convolutional and time-domain baselines by 0.26 dB to 0.63 dB SI-SDR under the remaining three conditions while achieving the strongest perceptual (DNSMOS) quality under dataset mismatch. The largest evaluated configuration, U-PAST-H (2.40M parameters), is consistently the strongest variant of the family, offering an attractive performance-to-cost trade-off at a small parameter footprint.
\end{abstract}


\section{Introduction}
\label{sec:intro}

Speech signals in real environments are often corrupted by background noise that significantly impacts speech quality and intelligibility, creating difficulties for human listeners and degrading downstream speech-processing systems.
Thus, recovery of clean speech from noisy environments is an active research problem. In the single-channel setting, due to the lack of spatial information, a robust modeling of both spectral structure and temporal context of speech and noise is essential.
Several approaches to single-channel speech enhancement have been proposed. Some methods operate in the time domain \cite{kong2022speech}, while others use spectral representations such as mel-frequency spectra \cite{shao2025cleanmel, yang2025mel} or the short-time Fourier transform (STFT) \cite{li2025dccrn}. Various model architectures have been proposed, including convolutional structures such as U-Nets \cite{nustede2021towards, nustede2023single}, attention-based models \cite{giri2019attention, yu2022dual}, and generative approaches \cite{richter2023speech, shetu2025gan}. 
Transformer-based models such as DPT-FSNet \cite{dang2022dpt} and TSTNN \cite{wang2021tstnn} show strong interpretability and temporal modeling capability but DPT-FSNet requires sub-band processing, which incurs high computational cost.
Vision Transformers (ViT) \cite{dosovitskiy2020vit} have demonstrated efficient capture of long-range dependencies and effective modeling of global context in images.
In \cite{bahmei2025real},  a ViT-based architecture is applied to speech enhancement by stacking time- and frequency-domain representations into a single image-like input. The approach aims to predict clean and noise masks for enhanced output spectrogram estimation via a transformer read-out, while applying the (noisy) input phase during inverse short-time Fourier transform reconstruction. Architecturally, \cite{bahmei2025real} is a transformer-only encoder–readout model rather than a hybrid transformer U-Net since it lacks a convolutional decoder with hierarchical upsampling and skip connections, and it does not process phase. Because phase information is crucial for perceptual quality, reusing the noisy input phase limits achievable enhancement quality. 

To address the above considerations, we propose U-PAST, a transformer-U-Net network for speech enhancement that operates directly on complex-valued spectrograms. We convert the magnitude-phase input spectra into multiple patch tokens. Similarly to how a transformer handles sequences, the self-attention layer allows each patch token to interact with all others, capturing relevant global spectro-temporal patterns. In contrast to \cite{bahmei2025real}, U-PAST pairs the ViT encoder with a complex-valued U-Net decoder that includes hierarchical upsampling and lateral skip connections, and that reconstructs both magnitude and phase rather than reusing the noisy input phase.

\begin{figure*}[t]
    \centering
    \includegraphics[width=0.9\textwidth]{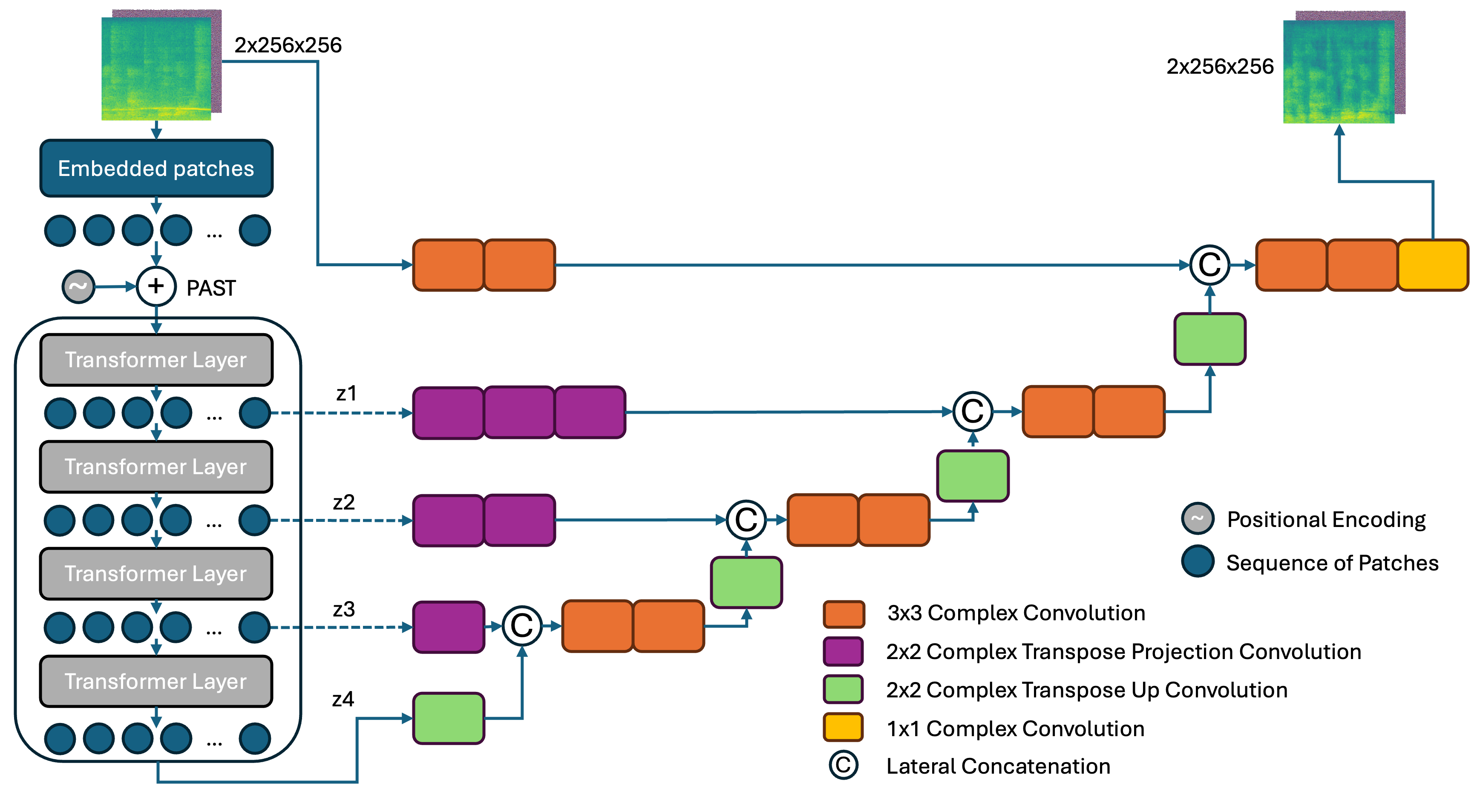}
    \caption{Hybrid U-PAST Architecture.
    The transformer module (left, depicted in the four layer configuration) encodes STFT magnitude and phase into token sequences. Green and orange blocks represent the CUNet decoder with hierarchical upsampling. Purple blocks denote feature projection layers that connect transformer outputs to the decoder. The yellow block is the prediction head for output feature reconstruction.}
    \label{fig:vitcvunet}
    \vspace*{-3mm}
\end{figure*}
\section{Related Work}
\label{sec:background}

\subsection{Audio Spectrogram Transformer}
\label{ssec:subvit}

The Audio Spectrogram Transformer (AST) \cite{gong21b_interspeech} adapts the ViT architecture from image processing to audio by representing a magnitude spectrogram as a sequence of tokens. Instead of using convolutional operations, the AST divides the spectrogram into patches and processes them using self-attention, enabling effective modeling of global context and long-range dependencies.

Given a spectrogram $x \in \mathbb{R}^{C \times F \times T}$, where $C$, $F$, and $T$ denote the number of channels, frequency bins, and time frames, respectively, the spectrogram is divided into $N$ patches of size $P \times P$. Each patch is then flattened and projected into an embedding space. The patch extraction and embedding (tokenization) process can be formulated as

\begin{equation}
z_0^i = x_p^i E + E_{\text{pos}}[i,:],
\quad
x_p^i = \mathrm{Flatten}(x_i) \in \mathbb{R}^{1 \times P^2 C},
\end{equation}

where $x_i$ denotes the $i$-th spectrogram patch, $i=1,\ldots,N$, and $N=FT/P^2$ (non-overlapping patches) or $N=\left(\lfloor(F-P)/S\rfloor+1\right)\left(\lfloor(T-P)/S\rfloor+1\right)$ (overlapping patches with stride $S$). A learnable [CLS] token, denoted by $x_p^0$, is prepended to the sequence. Similar to ViT, the input sequence contains $N+1$ tokens in total. The frequency and time dimensions, $F$ and $T$, are assumed to be divisible by the patch size $P$, ensuring that $N$ is an integer. The term $E \in \mathbb{R}^{P^2C \times D}$ denotes a learnable linear projection that maps each flattened patch to the embedding dimension $D$. To preserve positional information, a learnable positional encoding $E_{\text{pos}} \in \mathbb{R}^{(N+1) \times D}$ is added to the patch embeddings, where $E_{\text{pos}}[i,:] \in \mathbb{R}^{1 \times D}$ provides positional information for the $i$-th token.

All embedding tokens $z_0^i$ are passed passed through a stack of Transformer encoder layers. Through self-attention, the model can learn to assign different importance to different regions of the spectrogram, allowing the AST to effectively capture both local and global dependencies for audio classification.

\subsection{Complex-Valued Networks}
\label{ssec:cvunet}

Deep learning approaches to complex spectrogram processing fall broadly into two families: real-valued networks (RVNNs), which treat the real and imaginary (or magnitude and phase) components as separate real-valued channels, and complex-valued networks (CVNNs), which operate natively on complex arithmetic. RVNNs remain the more widely used choice in practice, but CVNNs, whose roots trace back to the complex least-mean-squares formulation of Widrow et al. \cite{widrow1975complex}, have attracted growing attention. They have since been explored well beyond audio, including for image denoising \cite{quan2021image}, object discovery \cite{lowe2022complex}, and medical imaging such as magnetic resonance imaging (MRI) \cite{dedmari2018complex}. Regardless of the numerical representation chosen, applying deep learning to complex spectrograms has driven substantial progress on perceptually oriented tasks such as phase retrieval \cite{griffin1984signal}, speech enhancement \cite{hu2020dccrn}, and speaker separation \cite{wang2021multi}, through advances spanning network architecture, training strategy, and loss function design. More recent efforts push in complementary directions, including low-latency streaming inference, universal models that generalize across enhancement tasks, and further gains in performance and efficiency, alongside growing interest in generative formulations of complex spectrogram reconstruction. A comprehensive survey of this literature is provided by Xie and Tan \cite{xie2025survey}.

In the audio domain specifically, complex-valued networks process complex STFT magnitude and phase information directly, rather than separating them into two streams or indirectly estimating a complex-valued output through, e.g., masking. Among them, complex-valued U-Nets \cite{nustede2024generalization} focus on single-channel speech enhancement, with the aim of improving noise suppression while preserving speech intelligibility in real-world conditions.
In the present work, we employ the decoder branch of a complex U-Net together with its lateral skip connections to obtain a high-quality audio reconstruction from the transformer model-encoded inputs.
\section{Methods}
\label{sec:methods}

\subsection{Complex-Valued AST Encoder}
\label{ssec:vitencoder}

The U-PAST architecture (cf.~Fig.~\ref{fig:vitcvunet}) adopts an AST-based encoder structure, which is derived from the ViT architecture originally developed for the image processing and widely used in the image domain such as medical image segmentation tasks \cite{hatamizadeh2021swin, hatamizadeh2022unetr}. It has also been proposed for audio classification, cf.\ section \ref{ssec:subvit}, where it is employed without a decoder for signal reconstruction. Motivated by these successes in vision and audio classification, we adapt it for speech enhancement by extending the approach to the complex domain and augmenting it with a convolutional decoder. A magnitude-phase STFT feature map of size $2 \times 256 \times 256$ is regarded as a 2D image with height $F$, width $T$ and channel dimension $C$. This input is partitioned into non-overlapping patches with a patch size of $16\times 16$. Each patch is then projected into an embedding space via a linear embedding layer, resulting in a sequence of $N = 16 \cdot 16 = 256$ tokens. A positional encoding vector is added to each token, indicating its position in the original $256 \times 256$ input feature map. 
Cf.~Algorithm~\ref{alg:patch} for a detailed description of the audio feature tokenizer.

To study the complexity–performance trade-off, we define several model ablations that vary the number of encoder layers and the input patch size, while holding the decoder depth, embedding (hidden) dimension, and number of attention heads fixed.
Table~\ref{tab:vit} lists the variants from the shallowest, U-PAST-S (1 layer), through U-PAST-B (4 layers), to the deeper U-PAST-L (8 layers) and U-PAST-H (12 layers) models, all using a hidden dimension of 96 and 3 attention heads.
Parameter numbers, MACs, and RTF for each model are reported in Table~\ref{tab:complexity}.

\begingroup
\scriptsize
\begin{algorithm}[t]
\caption{ViT-based complex features patch embedding}
\label{alg:patch}
\begin{algorithmic}[1]
\Require Input feature $X \in \mathbb{R}^{C \times F \times T}$, patch size $(P_f, P_t)$, embedding dimension $D$
\Ensure Patch embeddings $\mathcal{Z}$

\State Pre-allocate embedding matrix: $\mathcal{Z} \leftarrow \text{zeros}(N, D)$
\State Calculate number of patches: $N = \left\lfloor \frac{F}{P_f} \right\rfloor \cdot \left\lfloor \frac{T}{P_t} \right\rfloor$
\State Define learnable projection matrix: $E \in \mathbb{R}^{(C \cdot P_f P_t) \times D}$
\State Define 2D-sinusoidal positional encoding: $E_{\text{pos}} \in \mathbb{R}^{N \times D}$
\State Initialize token index: $i \leftarrow 0$

\For{$f = 0$ \textbf{to} $F - P_f$ \textbf{step} $P_f$}
    \For{$t = 0$ \textbf{to} $T - P_t$ \textbf{step} $P_t$}
        
        \State Extract patch: $p = X[:, \; f:f+P_f, \; t:t+P_t]$
        
        \State Flatten patch to row vector: $\hat{p} = \mathrm{Flatten}(p) \in \mathbb{R}^{1 \times (C \cdot P_f P_t)}$
        
        \State Project and add positional encoding: $z_i = \hat{p}E + E_{\text{pos}}[i,:]$ 
        
        \State Store patch embedding: $\mathcal{Z}[i,:] = z_i$
        
        \State Increment index: $i \leftarrow i + 1$
        
    \EndFor
\EndFor

\State \Return $\mathcal{Z}$
\end{algorithmic}
\end{algorithm}
\endgroup

\subsection{Convolutional U-Net Decoder}
\label{ssec:cvunetdec}

The decoder consists of four U-Net–style complex-valued upsampling layers, following the standard U-Net decoder architecture. Each upsampling block includes a $2 \times 2$ transposed convolution followed by two $3 \times 3$ convolutional blocks with a residual connection, depicted by green and orange blocks in Fig.~\ref{fig:vitcvunet}. Instead of using real-valued convolutions, we adopt complex-valued convolutional blocks \cite{nustede2024generalization}, each followed by a complex parametric rectified linear unit (CPReLU) \cite{pandey2019exploring} activation, complex batch normalization \cite{pandey2019exploring}, and a drop out layer with rate of 0.2. In the prediction head, a $1 \times 1$ complex-valued convolution layer is applied with an output channel size of 2, corresponding to magnitude and phase of the enhancement outputs.

To investigate the interplay of transformer encoder and U-Net decoder, we also compare them to two complex U-Net models (CUNet-S and CUNet-B) that use a four-layer U-Net encoder together with the decoder described above, cf.\ Section~\ref{ssec:trainconfigs} for implementation details.

\subsection{Feature Projection}
\label{ssec:midproj}

Since the transformer works with the token sequence of size $[B, N, D]$ while the U-Net decoder blocks expect 2D feature maps of size $[B, D, F, T]$, a projection function \cite{hatamizadeh2021swin} acts as a bridge that transforms tensor dimensionality according to 
\begin{equation*}
N \times D \xrightarrow{view} F_g \times {T_g} \times {D} \xrightarrow{permute} D \times {F_g} \times {T_g}.
\end{equation*}
Here, $F_g$ and $T_g$ represent the number of tokens along the frequency and time axes, respectively, $B$ denotes batch size, and $D$ the transformer's hidden dimension (Table \ref{tab:vit}).
To match the skip connections from the encoder to the decoder, we utilize $2\times2$ complex transposed projection convolutions as shown by the purple blocks in Fig.~\ref{fig:vitcvunet}. These layers take a transformer's hidden size $D$ as input channel and expand the spatial information by 2x, 4x, and 8x the base spatial patch grid size $(F_g, T_g)$, depending on the decoder-layer to which they project. Thus, depending on the target decoder stage, one to three such projection modules are stacked to reach the required spatial resolution.

\begin{table}[t]
\centering
\caption{Details of hybrid U-PAST encoder variants}
\label{tab:vit}
\begin{tabular}{l| c c c c}
\toprule
Model & Patch & Layers & Hidden & Heads \\
\midrule
U-PAST-S & 16x16 & 1 & 96 & 3 \\
U-PAST-B & 16x16 & 4 & 96 & 3 \\
U-PAST-L & 16x16 & 8 & 96 & 3 \\
U-PAST-H & 16x16 & 12 & 96 & 3 \\
\bottomrule
\end{tabular}
\end{table}

\subsection{Loss Function}
\label{ssec:lossf}
The training objective $\mathcal{L}$ is computed as the weighted sum of a reconstructed waveform loss $\mathcal{L}_{\text{Wave}}$, a multi-resolution STFT loss $\mathcal{L}_{\text{MR-STFT}}$, and a scale-invariant signal-to-distortion ratio (SI-SDR) \cite{le2019sdr} loss $\mathcal{L}_{\text{SI-SDR}}$:
\begin{equation}
\label{eq:wave}
\mathcal{L}_{\text{Wave}} = \| \hat{y} - y \|_1
\end{equation}
The multi-resolution STFT loss \cite{yamamoto2020parallel} averages a spectral convergence term $\mathcal{L}_{\text{SC}}$ and a log-magnitude term $\mathcal{L}_{\text{Mag}}$ over $M$ STFT resolutions, each parameterized by an FFT size, hop size, and window length:

\begin{equation}
\label{eq:sc}
\mathcal{L}_{\text{SC}} = \frac{\lambda_{sc}}{M}\sum_{m=1}^{M} \frac{\left\| \hat{Y}_{\text{Mag}}^{(m)} - Y_{\text{Mag}}^{(m)} \right\|_F}{\left\| Y_{\text{Mag}}^{(m)} \right\|_F}
\end{equation}

\begin{equation}
\label{eq:mrmag}
\mathcal{L}_{\text{Mag}} = \frac{\lambda_{mag}}{M}\sum_{m=1}^{M} \frac{1}{T_m F_m}\sum_{t,f} \left| \log Y_{\text{Mag}}^{(m)}(t,f) - \log \hat{Y}_{\text{Mag}}^{(m)}(t,f) \right|
\end{equation}

\begin{equation}
\label{eq:mrstft}
\mathcal{L}_{\text{MR-STFT}} = \mathcal{L}_{\text{SC}} + \mathcal{L}_{\text{Mag}}
\end{equation}

\begin{equation}
\label{eq:sisdr}
\mathcal{L}_{\text{SI-SDR}} = -\text{SI-SDR}(\hat{y}, y)
\end{equation}

\begin{equation}
\label{eq:loss}
\mathcal{L} = \lambda_1 \mathcal{L}_{\text{Wave}} + \lambda_2 \mathcal{L}_{\text{MR-STFT}} + \lambda_3 \mathcal{L}_{\text{SI-SDR}}
\end{equation}

Here, $y$ and $\hat{y}$ denote target and estimated waveform signals, and $Y_{\text{Mag}}^{(m)}$, $\hat{Y}_{\text{Mag}}^{(m)}$ denote the target and estimated magnitude spectrograms (of size $T_m \times F_m$) obtained via STFT at resolution $m$, with $\lVert \cdot \rVert_F$ the Frobenius norm. We use $M$=3 resolutions with FFT sizes $\{512, 1024, 2048\}$, hop sizes $\{50, 120, 240\}$, and window lengths $\{240, 600, 1200\}$, and set $\lambda_{sc}$=$\lambda_{mag}$=0.5. The overall weighting factors are $\lambda_1$=1.0, $\lambda_2$=1.0, and $\lambda_3$=0.05.
\section{Experimental Setup}
\label{sec:expsetup}

\subsection{Data}
\label{ssec:data}

Analyses are conducted on three corpora, the 2020 Microsoft Deep Noise Separation (DNS) Challenge dataset \cite{reddy2021interspeech}, the VoiceBank-DEMAND corpus \cite{botinhao2016investigating}, and the LibriMix corpus \cite{cosentino2020librimix}, all resampled to 16~kHz. The training set consists of 100 hours of clean audio from the DNS dataset, mixed with noise at SNRs sampled uniformly from 0 to 20 dB, with one hour of data set aside for validation.
To evaluate performance and robustness of trained models, we use four unseen test set conditions. The matched test condition uses 20 unseen speakers from DNS mixed with unseen noise types at SNRs from 0 to 15 dB. The acoustic mismatch condition adds reverberation to the same DNS test data. The remaining two conditions assess dataset mismatch, i.e.\ the models' ability to generalize across varying recording conditions and environments. The first uses the test portion of VoiceBank-DEMAND, which includes two speakers and five noise types.
The second is derived from the two-speaker LibriMix (Libri2Mix) test set \cite{cosentino2020librimix}, generated in \texttt{min} mode at 16~kHz from LibriSpeech \texttt{test-clean} \cite{panayotov2015librispeech}. As our models are trained for single-speaker enhancement, we retain only the first speaker (s1), a LibriSpeech utterance \cite{panayotov2015librispeech}, and mix it with WHAM! noise \cite{wichern2019wham}: the noisy s1 signal serves as the input and the clean s1 utterance as the target, with the second speaker discarded. The resulting test split contains 3,000 utterance pairs totaling 4~hours of audio, with per-utterance duration governed by \texttt{min} mode, i.e.\ each mixture is truncated to the length of its shorter constituent source. This condition probes generalization to speaker and noise characteristics that differ from both the DNS and VoiceBank-DEMAND data.

\subsection{Training and Evaluation Configuration}
\label{ssec:trainconfigs}

Inputs are random segments of length 25,500 samples (at 16~kHz) processed by an STFT with window-length 400 samples, window-shift 100 samples, and 512-point FFT.
Log-magnitude and phase of the STFT are calculated, and after removal of the DC component the resulting STFT feature is denoted as $I = \mathrm{STFT}(w) \in \mathbb{R}^{2 \times 256 \times 256}$. The first dimension corresponds to the magnitude and phase components (C), the second represents frequency bins (F), and the last dimension denotes time frames (T).

The Adam optimizer is used for training with an initial learning rate of $0.001$, applied to all trainable parameters. A cosine learning rate schedule with linear warmup is adopted to stabilize early training and improve convergence, where the number of warmup steps is set to 100 and the total number of training steps is defined as the product of number of epochs (n=500) and iterations per epoch.

Evaluation is conducted using SI-SDR, PESQ, and DNSMOS \cite{reddy2022dnsmos} measures to assess speech quality and perceptual metrics. To access computational complexity, we additionally report multiply-accumulate operations (MACs) and real-time factor (RTF). MACs are measured for a single forward pass over a fixed-duration dummy input (4-second at 16 kHz, chunked identically to inference) using ptflops, which hooks directly into individual modules and therefore accounts for recurrent layers, such as the LSTMs, that may not be properly decomposed by ATen-operator-based counters. RTF is defined as the average wall-clock inference time divided by the duration of the input audio, averaged over 20 runs following 3 warmup iterations; RTF < 1 indicates faster-than-real-time processing. All MACs and RTF (GPU) measurements are obtained on an NVIDIA A100-80G are reported alongside parameter counts in Table~\ref{tab:complexity}.

We further compare against complex-valued U-Net baselines that share our decoder but replace the transformer encoder with a standard convolutional encoder.
To this end, we implement two U-Nets (CUNet-S and CUNet-B) following the approach of \cite{nustede2024generalization} but reduced to 4-layers with consequently a parameter count that is in a similar range as for the U-PAST models. CUNet-S uses 32, 64, 128 and 256 channels in the four layers, whereas CUNet-B has channel dimensions 64, 128, 256 and 512.
Further baseline algorithms from the literature are MP-SENet \cite{lu2023mp}, a model with magnitude-phase processing and a tentatively small parameter count, and CleanUNet \cite{kong2022speech} that operates on time-domain inputs with an order of magnitude larger parameter count. We additionally include TF-GridNet \cite{wang2023tf}, a time-frequency domain network with a parameter count similar to the U-PAST models, as a further point of comparison for computational complexity (Table~\ref{tab:complexity}); as no official TF-GridNet implementation is publicly available, we use the TF-GridNet-based code released for X-TF-GridNet \cite{hao2024if}.

\begin{table}[ht]
\centering
\caption{Model complexity comparison.}
\label{tab:complexity}
\begin{tabular}{l|cccc}
\toprule
Method & Domain & Params$\downarrow$ (M) & MACs$\downarrow$ (G) & RTF (A100)$\downarrow$ \\
\midrule
U-PAST-S & T-F & 1.17 & 4.39 & 0.0126 \\
U-PAST-B   & T-F & 1.51 & 4.48 & 0.0132 \\
U-PAST-L  & T-F & 1.96 & 4.60 & 0.0142 \\
U-PAST-H  & T-F & 2.40 & 4.71 & 0.0151 \\
\midrule
CUNet-S \cite{nustede2024generalization}   & T-F & 2.1 & 8.7 & 0.0081 \\
CUNet-B \cite{nustede2024generalization}   & T-F & 8.34 & 34.53 & 0.0115 \\
TF-GridNet \cite{wang2023tf}        & T-F & 1.34 & 85.58 & 0.1880 \\
MP-SENet \cite{lu2023mp}        & T-F & 2.26 & 164.01 & 0.0523 \\
CleanUNet \cite{kong2022speech} & T & 46.07 & 17.22 & 0.0058 \\
\bottomrule
\end{tabular}
\end{table}

\begin{table*}[ht]
\centering
\caption{Enhancement results for matched DNS non-reverb condition. Training on DNS non-reverb data.}
\label{tab:matched_non_reverb_results}
\resizebox{0.98\linewidth}{!}{%
\begin{tabular}{l|cccc|cccc}
\toprule
Method
& PESQ-wb$\uparrow$ & PESQ-nb$\uparrow$ & STOI$\uparrow$ & SI-SDR$\uparrow$
& MOS-p808$\uparrow$ & MOS-sig$\uparrow$ & MOS-bak$\uparrow$ & MOS-ovr$\uparrow$ \\
\midrule
U-PAST-S & 2.2959& 2.8651& 0.9404& 14.5479& 3.7743& 3.4333& 3.8047& 3.0546\\ 
U-PAST-B & 2.3659& 2.9076& 0.9426& 15.1634& 3.7699& 3.4630& 3.7942& 3.0743\\ 
U-PAST-L & 2.4263& 2.9613& 0.9458& 15.4931& 3.8172& 3.4516& 3.9056& 3.1194\\ 
U-PAST-H & 3.4076& 2.9654& 0.9464& 15.6853& 3.8443& 3.4582& 3.9000& 3.1208\\ 
\midrule
CUNet-S \cite{nustede2024generalization}   & 2.2752 & 2.9285 & 0.9467 & 15.5515 & 3.8834 & 3.4435 & 3.8932 & 3.1080 \\
CUNet-B \cite{nustede2024generalization}   & 2.3503 & 3.0147 & 0.9507 & 16.3142 & 3.9054 & 3.4866 & 3.8994 & 3.1500 \\
TF-GridNet \cite{wang2023tf}    & 2.8915 & 3.3572 & 0.9667 & 18.1282 & 3.8852 & 3.5429 & 4.0311 & 3.2520 \\
MP-SENet \cite{lu2023mp} & 2.4780& 3.0342& 0.9504& 16.2360& 3.7965& 3.4916& 3.8468& 3.1169\\
CleanUNet \cite{kong2022speech} & 2.2843 & 2.8713 & 0.9523 & 15.5621 & 3.6448 & 3.4423 & 3.9900 & 3.1457 \\
\midrule
Unprocessed & 1.5822 & 2.1607 & 0.9152 & 9.0709 & 3.1557 & 3.3920 & 2.6181 & 2.4827 \\
\bottomrule
\end{tabular}
}
\end{table*}

\begin{table*}[ht]
\centering
\caption{Enhancement results for matched DNS reverb condition. Training on DNS non-reverb data.}
\label{tab:matched_reverb_results}
\resizebox{0.98\linewidth}{!}{%
\begin{tabular}{l|cccc|cccc}
\toprule
Method
& PESQ-wb$\uparrow$ & PESQ-nb$\uparrow$ & STOI$\uparrow$ & SI-SDR$\uparrow$
& MOS-p808$\uparrow$ & MOS-sig$\uparrow$ & MOS-bak$\uparrow$ & MOS-ovr$\uparrow$ \\
\midrule
U-PAST-S & 1.8479& 2.5330& 0.8481& 9.6582& 3.2369& 2.6872& 2.9910& 2.1554\\ 
U-PAST-B & 1.7115& 2.3966& 0.8279& 9.7190& 3.2075& 2.7808& 3.0992& 2.2308\\ 
U-PAST-L & 1.7479& 2.4335& 0.8316& 9.3665& 3.2464& 2.8041& 3.1886& 2.2619\\ 
U-PAST-H & 1.7188& 2.4149& 0.8314& 9.7649& 3.2601& 2.8074& 3.1638& 2.2605\\ 
\midrule
CUNet-S \cite{nustede2024generalization}    & 1.5807 & 2.1600 & 0.8275 & 8.6090 & 3.3219 & 2.8340 & 3.1222 & 2.2682 \\
CUNet-B \cite{nustede2024generalization}    & 1.5263 & 2.1164 & 0.8082 & 8.2435 & 3.3700 & 2.9148 & 3.0510 & 2.3065 \\
TF-GridNet \cite{wang2023tf}    & 1.7322 & 2.4194 & 0.8156 & 7.4466 & 3.3721 & 3.0931 & 3.6531 & 2.6296 \\
MP-SENet \cite{lu2023mp}                   & 1.7911& 2.4126& 0.8191& 8.1388& 3.3531& 2.9142& 3.4024& 2.4217\\
CleanUNet \cite{kong2022speech}            & 1.2074 & 1.4949 & 0.6486 & 3.5693 & 2.6865 & 2.1507 & 3.5748 & 1.9002 \\
\midrule
Unprocessed & 1.8218 & 2.5205 & 0.8662 & 9.0328 & 2.7384 & 1.7603 & 1.4973 & 1.3923 \\
\bottomrule
\end{tabular}
}
\end{table*}

\begin{table*}[ht]
\centering
\caption{Enhancement results for mismatched VoiceBank-DEMAND. Training on DNS non-reverb data.}
\label{tab:mismatched_voicebank_results}
\resizebox{0.98\linewidth}{!}{%
\begin{tabular}{l|cccc|cccc}
\toprule
Method
& PESQ-wb$\uparrow$ & PESQ-nb$\uparrow$ & STOI$\uparrow$ & SI-SDR$\uparrow$
& MOS-p808$\uparrow$ & MOS-sig$\uparrow$ & MOS-bak$\uparrow$ & MOS-ovr$\uparrow$ \\
\midrule
U-PAST-S & 2.3634& 3.1808& 0.9283& 15.5709& 3.4543& 3.3714& 3.7842& 2.9894\\ 
U-PAST-B & 2.4299& 3.2573& 0.9312& 16.0915& 3.4576& 3.4182& 3.8488& 3.0579\\ 
U-PAST-L & 2.4221& 3.2531& 0.9308& 16.4133& 3.4804& 3.3922& 3.8604& 3.0434\\ 
U-PAST-H & 2.4912& 3.3330& 0.9324& 16.7224& 3.4923& 3.3878& 3.8945& 3.0549\\ 
\midrule
CUNet-S \cite{nustede2024generalization}    & 2.3698 & 3.3702 & 0.9275 & 15.6256 & 3.4576 & 3.3113 & 3.8210 & 2.9571 \\
CUNet-B \cite{nustede2024generalization}    & 2.3882 & 3.4048 & 0.9298 & 17.0499 & 3.5098 & 3.3458 & 3.8148 & 2.9887 \\
TF-GridNet \cite{wang2023tf}    & 2.7322 & 3.5145 & 0.9459 & 18.3441 & 3.5008 & 3.4539 & 3.9470 & 3.1283 \\
MP-SENet \cite{lu2023mp} & 2.6851& 3.4329& 0.9306& 16.2526& 3.4802& 3.3938& 3.8247& 3.0171\\
CleanUNet \cite{kong2022speech} & 2.2797 & 3.1862 & 0.9343 & 13.5252 & 3.3107 & 3.2441 & 3.8410 & 2.8971 \\
\midrule
Unprocessed & 1.9709 & 2.8784 & 0.9210 & 8.4434 & 3.0861 & 3.3243 & 3.1107 & 2.6836 \\
\bottomrule
\end{tabular}
}
\end{table*}

\begin{table*}[ht]
\centering
\caption{Enhancement results for mismatched LibriMix. Training on DNS non-reverb data.}
\label{tab:mismatched_librimix_results}
\resizebox{0.98\linewidth}{!}{%
\begin{tabular}{l|cccc|cccc}
\toprule
Method
& PESQ-wb$\uparrow$ & PESQ-nb$\uparrow$ & STOI$\uparrow$ & SI-SDR$\uparrow$
& MOS-p808$\uparrow$ & MOS-sig$\uparrow$ & MOS-bak$\uparrow$ & MOS-ovr$\uparrow$ \\
\midrule
U-PAST-S & 1.5928& 2.2152& 0.8553& 9.6100& 3.2820& 3.1039& 3.3085& 2.5683\\ 
U-PAST-B & 1.6238& 2.2558& 0.8589& 9.8480& 3.2792& 3.1213& 3.4747& 2.6469\\ 
U-PAST-L & 1.6543& 2.2908& 0.8648& 10.0886& 3.3481& 3.1626& 3.5414& 2.7056\\ 
U-PAST-H & 1.6488& 2.3019& 0.8646& 10.2973& 3.3880& 3.1852& 3.5677& 2.7362\\ 
\midrule
CUNet-S \cite{nustede2024generalization}    & 1.6107 & 2.3142 & 0.8651 & 9.8565 & 3.4176 & 3.1390 & 3.6172 & 2.7251 \\
CUNet-B \cite{nustede2024generalization}    & 1.6508 & 2.3899 & 0.8710 & 10.4318 & 3.4394 & 3.2341 & 3.5463 & 2.7721 \\
TF-GridNet \cite{wang2023tf}    & 1.9594 & 2.6169 & 0.8933 & 12.0273 & 3.4575 & 3.3833 & 3.8429 & 3.0105 \\
MP-SENet \cite{lu2023mp} & 1.6525& 2.3128& 0.8601& 10.5615& 3.3544& 3.2251& 3.5306& 2.7395\\
CleanUNet \cite{kong2022speech}& 1.5334 & 2.0973 & 0.8704 & 10.3892 & 3.1212 & 3.1165 & 3.9125 & 2.8224 \\
\midrule
Unprocessed & 1.1588 & 1.6585 & 0.7959 & 3.4520 & 2.6254 & 2.4617 & 1.8122 & 1.7538 \\
\bottomrule
\end{tabular}
}
\end{table*}

\begin{figure*}[t]
    \centering
    \includegraphics[width=0.9\textwidth]{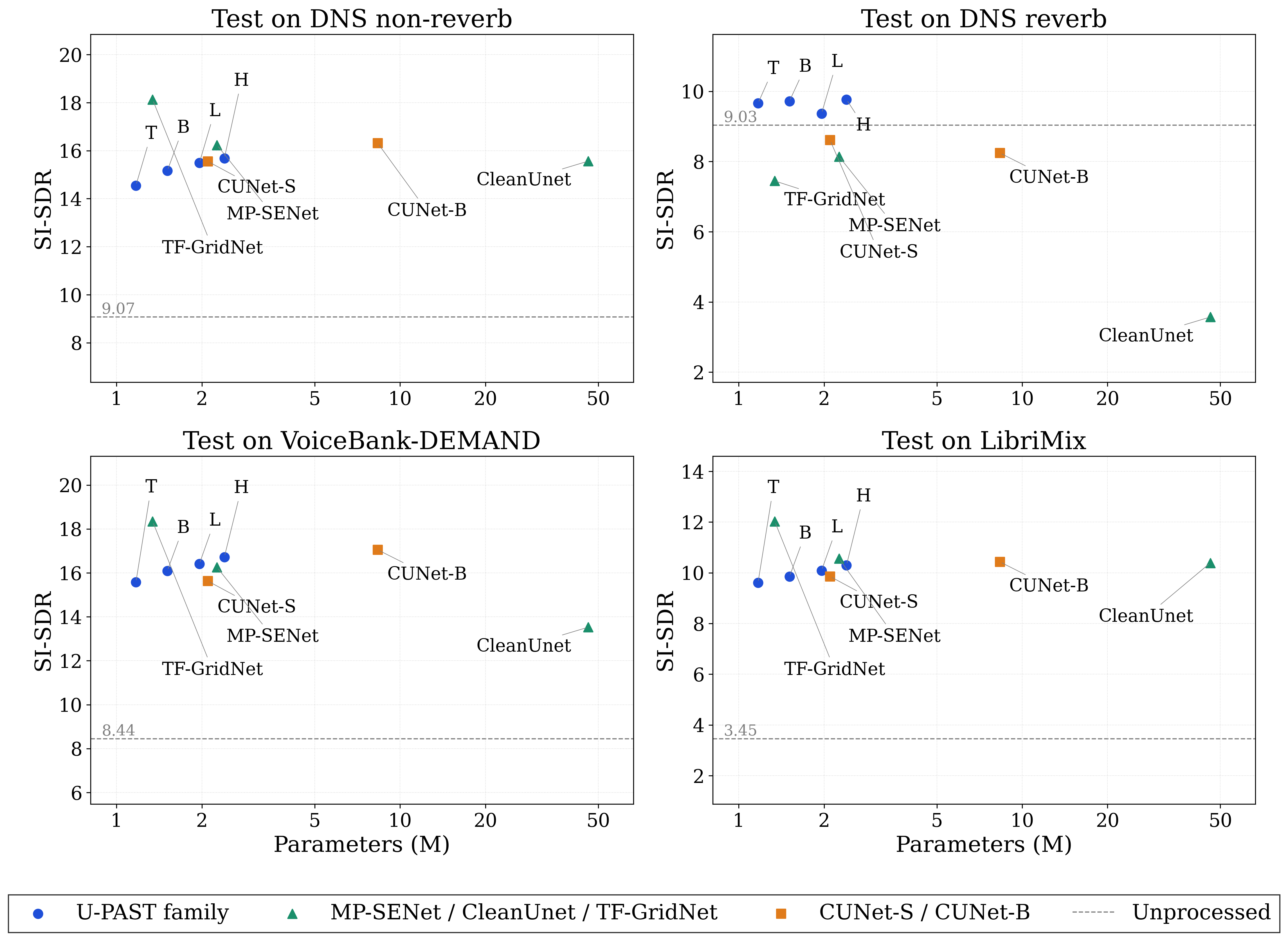}
    \caption{Speech-enhancement SI-SDR performance versus model size on four test conditions.}
    \label{fig:sisdrparams}
\end{figure*}

\begin{figure*}[t]
    \centering
    \includegraphics[width=0.9\textwidth]{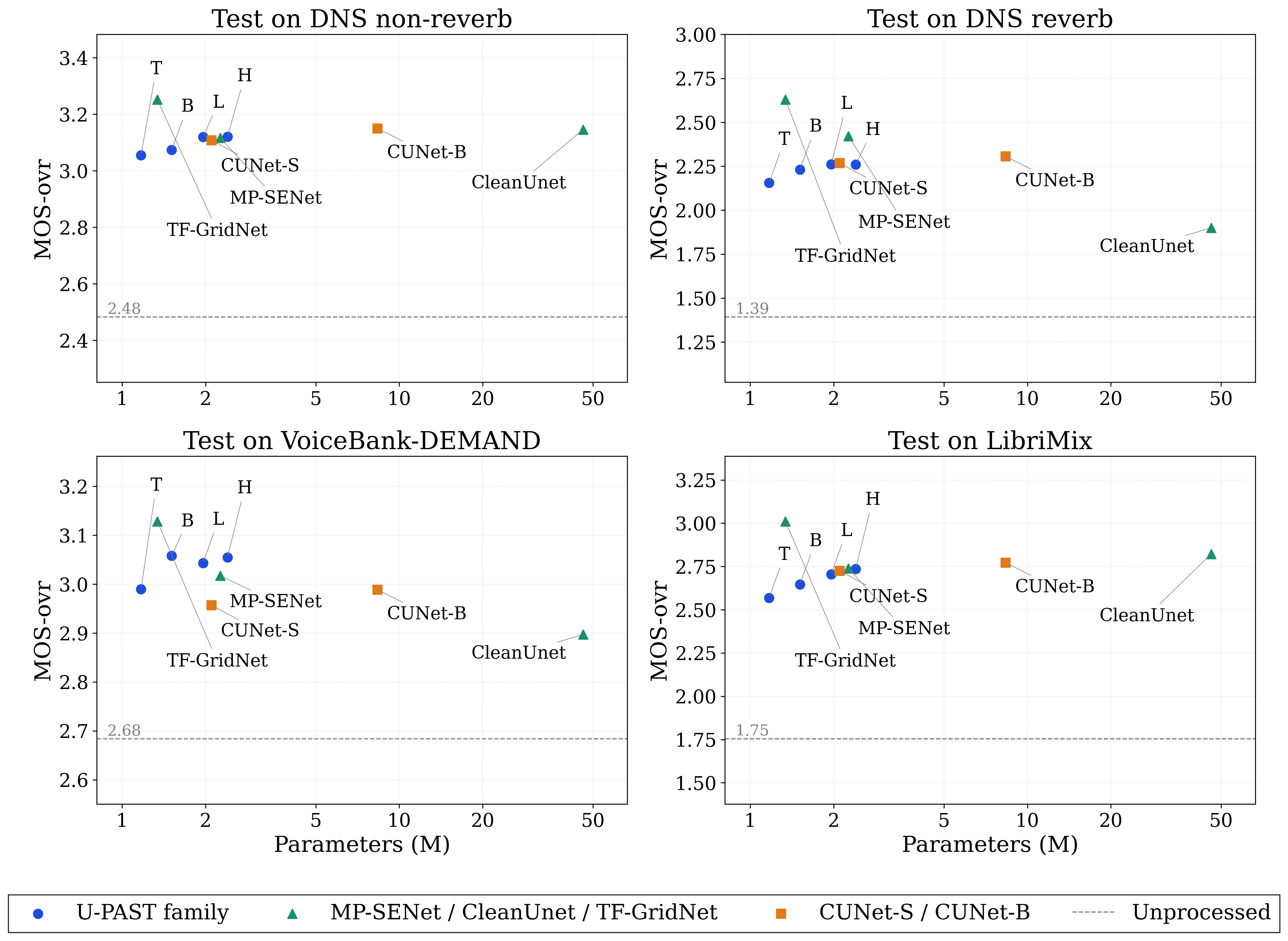}
    \caption{Speech-enhancement MOS-ovr performance versus model size on four test conditions.}
    \label{fig:mosparams}
\end{figure*}
\section{Results and Discussion}
\label{sec:results}

Table~\ref{tab:complexity} reports model complexity, and Tables~\ref{tab:matched_non_reverb_results}--\ref{tab:mismatched_librimix_results} report enhancement performance under the matched condition (DNS non-reverb), the acoustic mismatch condition (DNS reverb), and two dataset mismatch conditions (VoiceBank-DEMAND and LibriMix), respectively. Across all four test conditions, the U-PAST family occupies a markedly different point in the complexity-performance trade-off compared to the convolutional and time-domain baselines: it is among the smallest and least costly model classes evaluated, achieves the best SI-SDR of any evaluated model under acoustic mismatch (DNS reverb), and trails the strongest baseline, TF-GridNet, in three of the four conditions by 1.62–2.44~dB SI-SDR in the remaining three conditions, at a fraction of TF-GridNet's compute cost. Within the family, the largest evaluated configuration, U-PAST-H, is consistently the strongest variant.

\textbf{Complexity.} U-PAST-S is the smallest and least costly configuration evaluated, at 1.17M parameters and 4.39~GMAC, roughly $39\times$ fewer parameters than CleanUNet (46.07M), $7\times$ fewer than CUNet-B (8.34M), and about half the parameter count of MP-SENet (2.26M), while being similar in size to TF-GridNet (1.34M). U-PAST also requires $8\times$, $37\times$, and $18$--$19\times$ fewer MACs than CUNet-B, MP-SENet, and TF-GridNet, respectively. Both parameter count and MACs increase only mildly across the U-PAST family (1.17M--2.40M; 4.39--4.71~GMAC), in contrast to CUNet-B, which roughly quadruples the MACs of CUNet-S; MP-SENet, whose attention mechanism incurs by far the largest MAC count of all evaluated models (164.01~G); and TF-GridNet, whose MACs (85.58~G) are similarly disproportionate to its comparatively small parameter count. On GPU (A100), U-PAST's RTF (0.0126--0.0151) is faster than MP-SENet (0.0523) and more than an order of magnitude faster than TF-GridNet (0.1880), but remains higher than the purely convolutional CleanUNet (0.0058) and CUNet-S/B (0.0081/0.0115), indicating that the transformer encoder and complex-valued decoder are comparatively less hardware-efficient per MAC than standard convolutions.

\textbf{Matched condition (DNS non-reverb).} On this condition (Table~\ref{tab:matched_non_reverb_results}), TF-GridNet attains the best SI-SDR of any evaluated model (18.13~dB); the best U-PAST configuration, U-PAST-H, reaches 15.69~dB SI-SDR, trailing TF-GridNet by 2.44~dB, CUNet-B (16.31~dB) by 0.63~dB, and MP-SENet (16.24~dB) by 0.55~dB, while surpassing CleanUNet (15.56~dB, 19$\times$ more parameters) and matching CUNet-S (15.55~dB) at a comparable parameter budget (2.40M vs.\ 2.10M). DNSMOS scores in this condition are led by TF-GridNet, which attains the best MOS-sig (3.54), MOS-bak (4.03), and MOS-ovr (3.25) of all evaluated models; CUNet-B retains the best MOS-p808 (3.91). U-PAST-H attains the highest PESQ-wb of any evaluated model in this condition (3.41), exceeding the next-best TF-GridNet (2.89).

\textbf{Acoustic mismatch (DNS reverb).} Table~\ref{tab:matched_reverb_results} shows that reverberant mismatch is challenging for the baselines: CUNet-S (8.61~dB), CUNet-B (8.24~dB), MP-SENet (8.14~dB), and TF-GridNet (7.45~dB) all fall below the unprocessed input (9.03~dB) on SI-SDR, and CleanUNet fails severely (3.57~dB), indicating that models trained exclusively on non-reverberant data tend to introduce distortion under reverberation rather than removing it. In contrast, every evaluated U-PAST variant exceeds the unprocessed baseline, with U-PAST-H achieving the best SI-SDR of any evaluated model (9.76~dB), ahead of U-PAST-B (9.72~dB), U-PAST-S (9.66~dB), and U-PAST-L (9.37~dB). U-PAST-S additionally attains the best PESQ-wb (1.85) and PESQ-nb (2.53) of all models, the only configuration in this condition to improve on the unprocessed input's PESQ scores (1.82 / 2.52) rather than degrade them. Unprocessed audio retains the best STOI (0.866), though U-PAST-S is closest among processed outputs (0.848). DNSMOS scores in this condition are led by TF-GridNet, which attains the best MOS-p808 (3.37), MOS-sig (3.09), MOS-bak (3.65), and MOS-ovr (2.63) of all evaluated models, despite its comparatively weak SI-SDR.

\textbf{Dataset mismatch (VoiceBank-DEMAND).} On VoiceBank-DEMAND (Table~\ref{tab:mismatched_voicebank_results}), TF-GridNet attains the best SI-SDR overall (18.34~dB), ahead of CUNet-B (17.05~dB); U-PAST-H reaches 16.72~dB, trailing TF-GridNet by 1.62~dB and CUNet-B by 0.33~dB despite using $3.5\times$ fewer parameters than CUNet-B (2.40M vs.\ 8.34M), while still exceeding MP-SENet (16.25~dB) and CUNet-S (15.63~dB). TF-GridNet also leads most perceptual sub-scores in this condition, attaining the best PESQ-wb (2.73), PESQ-nb (3.51), STOI (0.946), MOS-sig (3.45), MOS-bak (3.95), and MOS-ovr (3.13) of all evaluated models; CUNet-B retains the best MOS-p808 (3.51), narrowly ahead of TF-GridNet (3.50). Among the remaining models, U-PAST attains competitive perceptual scores at a fraction of the parameter count of CUNet-B and CleanUNet: U-PAST-H's MOS-bak (3.89) is second only to TF-GridNet, and U-PAST-B attains the best MOS-sig (3.42) and MOS-ovr (3.06) among the non-TF-GridNet models.

\textbf{Further dataset mismatch (LibriMix).} On the LibriMix condition (Table~\ref{tab:mismatched_librimix_results}), TF-GridNet attains by far the best SI-SDR (12.03~dB), ahead of MP-SENet (10.56~dB), CUNet-B (10.43~dB), and CleanUNet (10.39~dB); the best U-PAST variant, U-PAST-H, reaches 10.30~dB, a gap of 1.73~dB behind TF-GridNet and 0.26~dB behind MP-SENet, ahead of CUNet-S (9.86~dB) and U-PAST-L (10.09~dB). TF-GridNet also leads most perceptual sub-scores, attaining the best PESQ-wb (1.96), PESQ-nb (2.62), STOI (0.893), MOS-p808 (3.46), MOS-sig (3.38), and MOS-ovr (3.01) of all evaluated models. CleanUNet retains the best MOS-bak (3.91).

\textbf{Effect of encoder size.} Across the evaluated encoder depths (U-PAST-S: 1 layer; U-PAST-B: 4 layers; U-PAST-L: 8 layers; U-PAST-H: 12 layers; cf.\ Table~\ref{tab:vit}), increasing transformer capacity yields a near-monotonic SI-SDR improvement in three of the four conditions (matched, VoiceBank-DEMAND, LibriMix), where SI-SDR rises consistently from U-PAST-S to U-PAST-H. The only exception is the acoustic mismatch (reverb) condition, where U-PAST-L (9.37~dB) dips slightly below U-PAST-B (9.72~dB) before U-PAST-H recovers to the best result of any evaluated model (9.76~dB). U-PAST-H is thus the strongest configuration of the family in every test condition, at a modest additional cost in parameters and MACs relative to U-PAST-S (2.40M/4.71~GMAC vs.\ 1.17M/4.39~GMAC). Fig.~\ref{fig:sisdrparams} and Fig.~\ref{fig:mosparams} illustrate this trade-off, situating the U-PAST family's SI-SDR and MOS-ovr against parameter count relative to the baselines.

Overall, U-PAST attains the best SI-SDR of any evaluated model under acoustic mismatch (DNS reverb), and trails the strongest baseline, TF-GridNet, in three of the four conditions by 1.62--2.44~dB SI-SDR elsewhere, while requiring $18$--$19\times$ fewer MACs and more than an order of magnitude lower RTF. Among the remaining, non-TF-GridNet baselines, U-PAST remains competitive or leading on several perceptual DNSMOS sub-scores, e.g.\ under dataset mismatch (VoiceBank-DEMAND), with only 1.17M--2.40M parameters, a fraction of the parameter count of CUNet-B (8.34M) and especially CleanUNet (46.07M).

\section{Conclusion}
\label{sec:conclusion}

In this work, we have introduced U-PAST, a phase-aware audio spectrogram transformer that, to our knowledge, constitutes the first hybrid transformer-U-Net for audio signal enhancement and the first to operate in the complex domain. The model encodes magnitude and phase from speech-in-noise spectrograms into token vectors, processes them with a multi-layer transformer, and reconstructs the enhanced speech directly via a complex-valued U-Net decoder.

Across matched, acoustic mismatch, and two dataset mismatch conditions, U-PAST variants spanning only 1.17M to 2.40M parameters remain competitive with convolutional and time-domain baselines that are up to $39\times$ larger (CleanUNet, 46.07M parameters), and are comparable in parameter count to the substantially stronger TF-GridNet baseline (1.34M parameters) despite requiring $18$--$19\times$ fewer MACs. U-PAST achieves the best SI-SDR of any evaluated model under acoustic mismatch (DNS reverb), where the largest configuration, U-PAST-H, reaches 9.76~dB, ahead of all baselines and the unprocessed input, while every other baseline, including TF-GridNet, degrades below the unprocessed input on this condition; in the remaining three conditions, U-PAST trails the strongest baseline, TF-GridNet, by 1.62--2.44~dB SI-SDR. U-PAST additionally attains the best PESQ-wb of any evaluated model under the matched condition (U-PAST-H, 3.41) and the best PESQ-wb/PESQ-nb under acoustic mismatch (U-PAST-S); its DNSMOS sub-scores remain competitive with the convolutional baselines at a fraction of their parameter count, though TF-GridNet attains the strongest DNSMOS sub-scores overall in three of the four conditions. Increasing encoder depth from 1 to 12 layers (U-PAST-S to U-PAST-H) yields a near-monotonic improvement in SI-SDR across three of the four conditions, with U-PAST-H the strongest configuration of the family in every condition evaluated.

This favorable accuracy-to-compute trade-off is not mirrored in wall-clock latency: on GPU (A100), U-PAST's RTF (0.0126--0.0151) is faster than MP-SENet (0.0523) and more than an order of magnitude faster than TF-GridNet (0.1880), but remains higher than the purely convolutional CleanUNet (0.0058) and CUNet baselines (0.0081--0.0115), indicating that the transformer encoder and complex-valued decoder are comparatively less hardware-efficient per MAC than standard convolutions. U-PAST thus offers an attractive footprint in terms of parameters and compute, but realizing this as a latency advantage will require further inference-side optimization.

Future work will explore smaller patch sizes in U-PAST configurations and target the inference-time efficiency of the architecture to translate its small footprint into a wall-clock advantage.

\bibliographystyle{unsrt}  
\bibliography{references}  






\end{document}